# Simplified Variable-Scaled Min Sum LDPC decoder for irregular LDPC Codes

Ahmed A. Emran [#1], Maha Elsabrouty [#2]

[#] Electronics and communications engineering department, E-JUST University
Borg El Arab, Alexandria, Egypt
[1] ahmed.emran@ejust.edu.eg
[2] maha.elsabrouty@ejust.edu.eg

*Abstract*—**Min-Sum decoding is widely used for decoding LDPC codes in many modern digital video broadcasting decoding due to its relative low complexity and robustness against quantization error. However, the suboptimal performance of the Min-Sum affects the integrated performance of wireless receivers. In this paper, we present the idea of adapting the scaling factor of the Min-Sum decoder with iterations through a simple approximation. For the ease of implementation the scaling factor can be changed in a staircase fashion. The stair step is designed to optimize the decoder performance and the required storage for its different values. The variable scaling factor proposed algorithm produces a non-trivial improvement of the performance of the Min-Sum decoding as verified by simulation results.**

## I. INTRODUCTION

Low-density parity check (LDPC) codes were first presented by Gallager [1] in the early 1960s. It has been shown that these codes have remarkable performance that is very close to Shannon limit when using iterative decoding. They become strong competitors to turbo codes [2] for error control for many digital communication systems. Because of their improved Bit Error Rate (BER) performance, these codes are used in many modern video broadcasting standards like DVB-S2 [4], DVB-T2 [5] [6], DVB-C2 [7], DTMB [8] and CMMB [9].

Usage of LDPC codes in video broadcasting standards assure better protection against errors which decrease the video quality. It also allows more data to be transported over a given channel [6]. The used decoding technique of LDPC code is an important parameter in its performance and its implementation complexity.

There are many types of decoding algorithms that can be used to decode LDPC codes. The soft-decision decoding algorithms are widely employed because of their superior performance over hard-decision algorithms. The log-likelihood ratio sum-product algorithms (LLR-SPA), developed by Mackay and Neal [10], are proven to achieve excellent capacity performance, by approaching to Shannon bound. However, one drawback for the LLR-SPA is the high complexity that implies large decoding delay that may be critical for some delay sensitive applications such as DVB. So, many modified approximations of LLR-SPA are developed to reduce its high complexity. One of the most important algorithms that satisfy this goal is the Min-Sum algorithm, Min-Sum is introduced in [11] as a simplification of LLR-SPA by using minimum operation instead of complex implemented tanh and $\tanh^{-1}$ functions. Many modified versions of Min-Sum algorithm were proposed to increase its performance with acceptable increasing in decoding complexity [12]-[17].

One of the most important modification is Scaled Min-Sum [12]-[14]. It is a modification of Min-Sum algorithm, where a scaling factor is used to decrease the error introduced by using the minimum operation. Scaled Min-Sum has a very good performance in regular LDPC codes. On the other hand, irregular LDPC codes require different scaling factor strategy [15] [16].

In [15], the scaling factor is calculated by approximating a nonlinear post-processing function to linear function. The non-linear function is highly affected by SNR and requires updating per iteration. In other words, irregular LDPC codes require different scaling factor per-iteration to achieve the optimum scaling scenario. Although changing the scaling factor with iterations gives a very good performance (low BER and avoiding error floor), it requires complex calculations (in design phase) and extra storage to store the scaling factor sequence.

In [16], two-dimension normalization was proposed where different scaling factor is used for each variable and check node degree. So two scaling vectors (α and β) are required for both check nodes' output and variable nodes' output respectively. Scaling factor vectors (α and β) calculation requires multi-dimension optimization (in design phase). In addition to design complexity, 2-D scaling factor implementation requires extra storage to store the scaling factors and extra complexity in scaling stage to choose different scaling factor for each degree.

Another modification of min-sum algorithm is selective max-min algorithm [17]. It uses maximum operation instead of summation in variable nodes processors. So it has lower complexity than scaled min-sum, however it has lower performance than scaled min-sum. In other words, selective max-min algorithm has an intermediate complexity and performance between scaled min-sum and min-sum algorithms.

In this paper, we propose a Simplified Variable Scaling (SVS) Min-Sum algorithm. SVS Min-Sum algorithm is based on using logical heuristic equation to calculate an easy implemented scaling factor sequence. This heuristic equation comes from observing the behavior of scaling factor sequence in [15], where the scaling factors increase exponentially with iterations and its final value equals 1. So we get the advantage of adaptive scaling factor with iteration as in [15] but with lower complexity in both design and implementation phases. In addition to lower implementation complexity, we avoid using two scaling stages for both variable and check nodes' output as in [16].

Simulation results of the proposed algorithm indicate that SVS Min-Sum algorithm can outperform Min-Sum Algorithm by 0.85 dB, and can outperform Scaled Min-Sum algorithm by 0.43 dB when applied on DVB-T2 LDPC codes with almost no extra complexity cost. The proposed algorithm requires only 0.05→0.2 dB more than LLR-SPA with much lower complexity. This performance enhancement can increase the quality of video reception over bad condition channels with little extra complexity.

The rest of the paper is organized as follows: Section II presents the necessary background on the SPA, Min-Sum, Scaled Min-Sum and Variable Scaled Min-Sum algorithms. Section III presents the SVS Min-Sum algorithm. The simulation results are displayed and discussed in Section IV. Finally, the paper is concluded in section V.

## II. REVIEW OF THE SPA AND MIN-SUM ALGORITHMS

An $(N, K)$ LDPC code is a binary code characterized by a sparse parity check matrix $\mathbf{H}_{M \times N}$ where M = N - K which can be represented by a tanner graph of variable nodes $n \in \{1 \cdots N\}$ and check nodes $m \in \{1 \cdots M\}$. We denote the set of variable nodes connected to a certain check node $m$ as $\mathcal{N}\{m\}$. A variable node $n$ is connected to the check node $m$ if $n \in \mathcal{N}\{m\}$. Furthermore, the set $\mathcal{N}\{m\} \setminus n$ denotes the set of variable nodes connected to the check node $m$ excluding $n$. Similarly, the set of check nodes connected to a certain variable node $n$ is denoted by $\mathcal{M}\{n\}$. A check node $m$ is connected to the variable node $n$ if $m \in \mathcal{M}\{n\}$. The set $\mathcal{M}\{n\} \setminus m$ denotes the set of check nodes connected to the variable node $n$ excluding $m$.

The main idea behind all belief propagation based algorithms is processing the received symbols iteratively in concatenated steps that can be seen over the Tanner graph as horizontal step followed by vertical step to improve the reliability of each decoded code symbol. The computed reliability measures of the code symbols at the end of any decoding iteration are used as inputs of the next iteration. This decoding iteration algorithm continues until a certain stopping criterion is satisfied.

To illustrate this concept consider: the reliability of a decoded symbol is measured by a *posteriori* probability $P(x_n | Y)$ for $1 \leq n \leq N$. Then the log-likelihood ratio LLR of each code bit is given by:

$$L(x_n) = \log \frac{P(x_n = 0 | Y)}{P(x_n = 1 | Y)} \qquad (1)$$

In each iteration, a message $r_{m \to n}$ is calculated in the horizontal step at each check node $m$ and is passed to all variable nodes $n$ if $n \in \mathcal{N}\{m\}$. Similarly each variable node $n$ sends a message $q_{n \to m}$ in the vertical step to all check nodes $m$ if $m \in \mathcal{N}\{n\}$.

The codeword is denoted by $X = [x_1, x_2, \ldots x_N]$ where $x_n \in \{0,1\}$. The LLR values of the corresponding received vector are denoted by $Y = [y_1, y_2, \ldots y_N]$.

In order to present SVS Min-Sum algorithm, we need to review the required background theory of the SPA, Min-Sum, Scaled Min-Sum and Variable Scaled Min-Sum algorithms. In the following subsections present a brief overview of these algorithms.

### A. Sum-Product algorithm (SPA)

The tanh-based SPA can be described in the following steps.

*1) Initialization step*
The initial values of the LLR can be obtained from the QAM demodulator output $y_n$. These initial values are used as $q_{n \to m}$, the first iteration's input message to the check node update step (Horizontal step).

*2) Horizontal step*
The horizontal step at a check node $m$ is dedicated to process the messages which are coming from the variable nodes $q_{n \to m}$ to calculate the reply messages $r_{m \to n}$ for all $n \in \mathcal{N}\{m\}$. So for each check node $m$

$$r_{m \to n} = \left( \prod_{n' \in N(m) \setminus n} sign(q_{n' \to m}) \right) \times 2 \tanh^{-1} \left( \prod_{n' \in N(m) \setminus n} \tanh\left( \frac{|q_{n' \to m}|}{2} \right) \right) \qquad (2)$$

*3) Vertical step*
The vertical step at a variable node $n$ is dedicated to process the messages which are coming from the check nodes $r_{m \to n}$ to calculate the reply messages $q_{n \to m}$ for all $m \in N\{n\}$. So for each variable node $n$ compute:

$$q_{n \to m} = y_n + \sum_{m' \in M(n) \setminus m} r_{m' \to n}(x_n) \qquad (3)$$

*4) Decision step:*
For each variable node, the LLR values are updated according to:

$$L(x_n) = y_n + \sum_{m \in M(n)} r_{m \to n}(x_n) \quad (4)$$

The LLR values are applied to hard decision to decide on the possible value of $x_n$ to be 1 if $L(x_n) < 0$ and zero otherwise. The syndrome is then calculated and checked to decide successful decoding if the syndrome is zero or to proceed to the next iteration if the syndrome condition is not satisfied. This process continues till either the code word is successfully decoded or the maximum iterations are exhausted.

Despite the optimum performance of the tanh-based SPA algorithm, it is difficult to implement due to the need to calculate $\tanh^{-1}(.)$ and $\tanh(.)$ functions which requires a series computation or saving in look up tables.

The tanh rule can be alternatively approximated using the Jacobi rule. This approximation yields the Min-Sum algorithm [11] which is more implementation friendly.

### B. Min-Sum algorithm

The Min-Sum algorithm follows the same steps as the tanh-rule SPA. It is composed of the same steps with only single change in the calculation of the horizontal step which can be manipulated to be: [11]

$$r_{m \to n} = \left( \prod_{n' \in \mathcal{N}(m) \setminus n} sign(q_{n' \to m}) \right) \times \min_{n' \in \mathcal{N}(m) \setminus n} \left( |q_{n' \to m}| \right) \quad (5)$$

The above algorithm is easier to implement as it gets rid of the tanh calculation. However, the approximation to the exponential calculations to the min (.) leads to some loss of performance compared to the tanh-based SPA algorithm. This loss of performance is partially recovered by Scaled Min-Sum algorithm.

### C. Scaled Min-Sum algorithm

In order to improve the performance of the Min-Sum algorithm, and make it closer to the performance of the tanh-based SPA algorithm, a constant scaling factor ($\alpha < 1$) can be applied to the check node update equation (Horizontal step) in all iterations. In other words, converts the Horizontal step to:

$$r_{m \to n} = \alpha \times \left( \prod_{n' \in \mathcal{N}(m) \setminus n} sign(q_{n' \to m}) \right) \times \min_{n' \in \mathcal{N}(m) \setminus n} \left( |q_{n' \to m}| \right) \quad (6)$$

This scaling factor can be calculated by either density evolution [13] or EXIT chart [14] to maximize the performance of Scaled Min-Sum algorithm.

### III. PROPOSED LOW COMPLEXITY VARIABLE SCALED MIN-SUM ALGORITHM

The Scaled Min-Sum algorithm gives very good performance for regular LDPC codes. On the other hand, Scaled Min-Sum performance for irregular LPDC codes is not good enough [15] [16]. In irregular codes, unequal message densities are sent from variable nodes with different degree. Unequal message densities require unequal scaling factor per iteration.

The idea of changing the scaling factor with the iteration for irregular LDPC codes is presented as a possible extension of the work in [16] and has been implemented in [15]. In [15], the calculation method of the required scaling factors is based on approximating a nonlinear converting function which converts the minimum operation's output back to LLR [15]. Despite of performance enhancement of this Variable scaled Min-Sum algorithm, it requires extra storage because we need different scaling factor values per iteration, associated with different degrees of the check nodes. The general fractional values taken by the scaling factor makes the multiplication by the update complex in implementation. The proposed Simplified Variable Scaled (SVS) Min-Sum algorithm addresses the particular point of simplified per-iteration updated scaling rule.

As stated in [15], the scaling factor should increase exponentially with iterations and its final value is 1. So we approximate the scaling factor graph to a stair graph with constant horizontal step S, and the scaling factor takes values which is exponential and at the same time easy to implement. The variable scaling factor can be calculated as:

$$\alpha = 1 - 2^{-\lceil i/S \rceil} \quad (7)$$

Where $\lceil i/s \rceil$ is the first integer larger than $i/S$. $i$ is the iteration index which take values {1, 2, 3, …}. So the scaling factor sequence is {0.5, 0.75, 0.875, 0.9375….}. And this sequence is:-

1) Easy to design, because it is based only on parameter S.
2) Does not need to store a specific scaling sequence for each code rate. The scaling sequence of every code rate require storing only the step size S and number of required shifts.
3) Easy to implement, because it only requires shifting right by $\lceil i/s \rceil$ then subtraction. Number of required shifts can be stored in a register and increased by 1 every S iterations.

A comparison between simulation results of proposed SVS Min-Sum algorithm and the other algorithms are displayed in section IV.

### IV. SIMULATION ENVIRONMENT AND RESULTS

The simulations are performed on the eIRA LDPC codes used in DVB-T2 standard [5] [6]. In DVB-T2 standard, there are two lengths of LDPC codes, normal length (N=64800) and short length (N=16200). For each length, there are many rates. The coding rates used for producing the simulation results are $R = 1/4$, $R = 1/2$ and $R = 3/4$ for short length codes and $R = 1/2$ and $R = 3/4$ for normal length codes. The other coding rates can also be decoded using the same scheme after finding the optimum $S$.

The data are produced as binary bits modulated using the challenging 256-QAM modulation scheme and sent over AWGN channel. The simulations are performed using MATLAB platform. The maximum number of iterations is set to 50 iterations.

The optimum (fixed) scaling factor ($\alpha$) of Scaled Min-Sum and optimum step size ($S$) of SVS Min-Sum are shown in table I. The optimum values of $\alpha$ and $S$ are calculated by simulation of the system with all possible values of scaling factor (or step size) and find ($\alpha$ or $S$) with minimum BER. These calculations are repeated for different SNR values and the optimum $\alpha$ (or $S$) gives BER<$10^{-6}$ while other values do not.

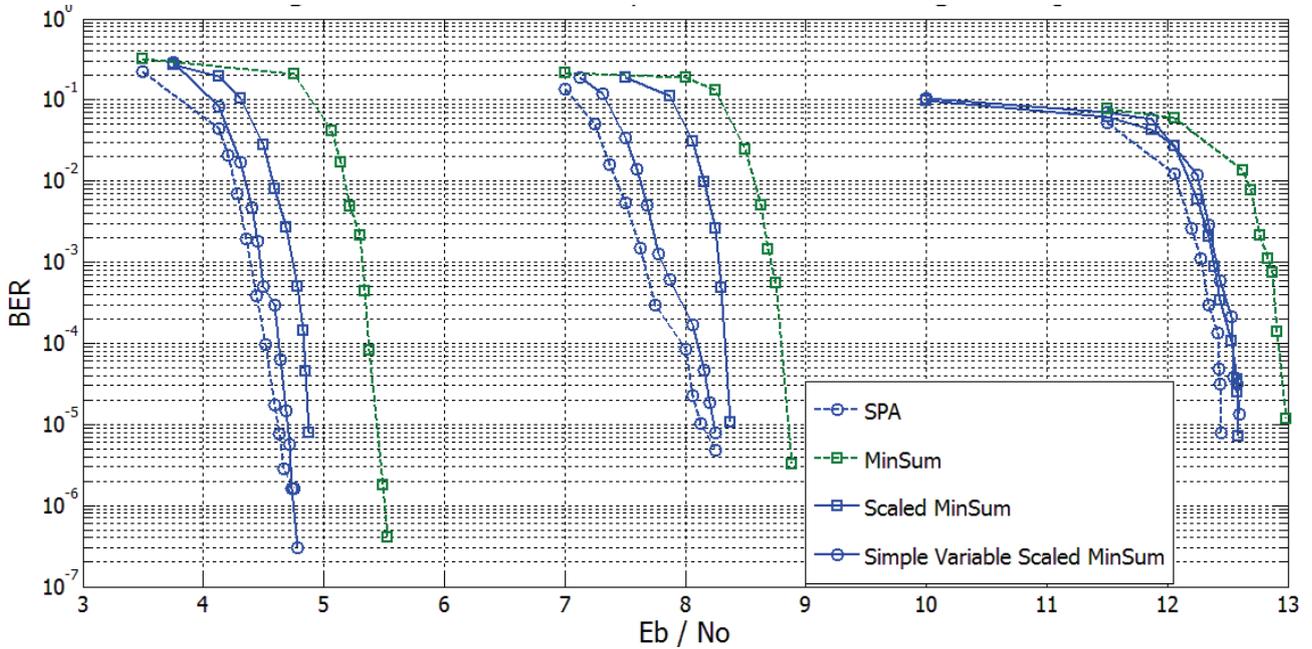

Fig.1. Short LDPC codes with rates = {0.25, 0.5, 0.75}

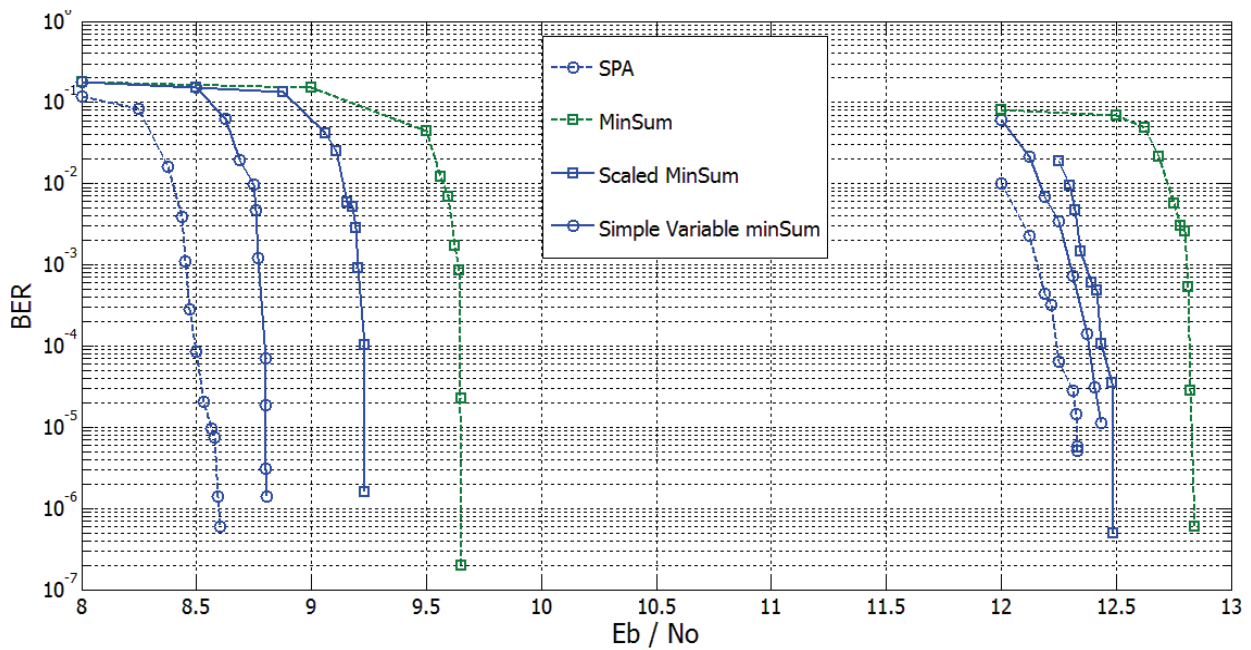

Fig.2. Normal length LDPC codes with rates = {0.5, 0.75}

TABLE I

USED CONSTANT α AND THE STEP SIZE

|  | Optimum constant α obtained through simulations | optimum step size $S$ obtained through simulations |
|---|---|---|
| Short 0.25 | 0.9375 | 7 |
| Short 0.50 | 0.9375 | 10 |
| Short 0.75 | 0.8125 | 13 |
| Normal 0.50 | 0.9375 | 7 |
| Normal 0.75 | 0.875 | 13 |

In fig 1, we present the BER of the DVB-T2 LDPC codes with short length for rates = 0.25 (most left), 0.5 (middle plots) and 0.75 (most right). Similarly, fig 2 illustrates the BER of the DVB-T2 LDPC codes with normal length for rates = 0.5 (at the left) and 0.75 (at the right). For both short and normal length codes, maximum number of iterations = 50.

As shown in figures 1 and 2, SVS Min-Sum algorithm has superior performance than constant Scaled Min-Sum algorithm with optimum α. SVS Min-Sum performance is very similar to SPA performance but with much lower complexity.

TABLE II

$E_b/N_o$ in dB TO ACHIEVE BER<$10^{-5}$

|  | SPA | SVS Min-Sum | Scaled Min-Sum | Min-Sum |
|---|---|---|---|---|
| Short 0.25 | 4.62 | 4.70 | 4.87 | 5.44 |
| Short 0.5 | 8.13 | 8.24 | 8.37 | 8.85 |
| Short 0.75 | 12.45 | 12.58 | 12.58 | 12.99 |
| Normal 0.5 | 8.56 | 8.80 | 9.23 | 9.65 |
| Normal 0.75 | 12.33 | 12.44 | 12.48 | 12.83 |

To be more accurate, table II illustrates the required $E_b/N_0$ in dB to achieve $BER \leq 10^{-5}$ for SPA, SVS Min-Sum, Scaled Min-Sum and Min-Sum algorithms.

As shown in table II, the SVS Min-Sum algorithm performance is very near to the SPA performance. In table III, differences between $E_b/N_0$ of SVS Min-Sum and other algorithms are calculated to display quantitatively the enhancement of performance by using the SVS Min-Sum algorithm.

TABLE III

COMPARING SVS MIN-SUM WITH OTHER ALGORITHMS

|  | Performance of SVS Min-Sum with respect to | | |
|---|---|---|---|
|  | SPA | Scaled Min-Sum | Min-Sum |
| Short 0.25 | -0.08 | 0.17 | 0.74 |
| Short 0.5 | -0.11 | 0.13 | 0.61 |
| Short 0.75 | -0.13 | 0 | 0.41 |
| Normal 0.5 | -0.24 | 0.43 | 0.85 |
| Normal 0.75 | -0.11 | 0.04 | 0.39 |

From the above results, it is clear that the SVS Min-Sum algorithm can provide better results that Min-Sum algorithm with improvement in the performance by 0.41→0.85 dB and better than the Scaled Min-Sum algorithm (with optimum α) with improvement in the performance by 0 to 0.43 dB. The SVS only requires 0.08 to 0.24 dB more than SPA with much lower complexity.

The simulation results assert the validity of the idea of changing or rather adapting the scaling factor of the Scaled Min-Sum and the simplicity of proposed adaptation method.

V. CONCLUSION AND FUTURE WORK

This paper proposed the idea of Simple Variable Scaled (SVS) Min-Sum decoder. The proposed algorithm examines the effect of simple per-iteration exponential update of the of the Scaled Min-Sum scaling factor for the study case of the DVB-T2 LDPC decoder. Simulation results indicated the validity of the idea of the heuristic easy to implement update of the scaling factor and its superior performance compared to both the regular Min-Sum and the fixed scaling factor Scaled Min-Sum algorithms. As for future work, the results can be extended to other irregular LDPC code.

ACKNOWLEDGMENT

This work is funded by the Egypt- Japan University of Science and Technology (E-JUST), Minister of High Education (MoHE)

This work has been supported by the Egyptian National Telecommunications Regularization Authority (NTRA) as part of the project "Design and implementation of DVB-T/T2 solution".